\documentclass[%
reprint,
superscriptaddress,
amsmath,amssymb,
aps,
prb,
]{revtex4-2}

\usepackage[T2A]{fontenc}    
\usepackage[utf8]{inputenc}

\usepackage{graphicx}
\usepackage[english]{babel} 
\usepackage{amssymb}
\usepackage{amsmath}
\usepackage{txfonts}
\usepackage{mathdots}
\usepackage[classicReIm]{kpfonts}
\usepackage[dvipsnames]{xcolor}
\usepackage{dcolumn}
\usepackage{bm}
\usepackage{hyperref}

\begin{document}

\preprint{APS/123-QED}

\title{Optical Cooling of Nuclear Spins in a CdTe/CdZnTe Quantum Well: The Impact of  Kinetic Local Fields on Cooling Efficiency.}

\author{V.M.Litvyak}
\affiliation{Spin Optics Laboratory, St-Petersburg State University, St-Petersburg, 198504, Russia}%
\author{P.S.Bazhin}
 \affiliation{Spin Optics Laboratory, St-Petersburg State University, St-Petersburg, 198504, Russia}
\author{R. Andr\'e}
\affiliation{Universit\'e Grenoble Alpes, CNRS, Institut N\'eel, 38000 Grenoble, France}
\author{K.V.Kavokin}%
\affiliation{Spin Optics Laboratory, St-Petersburg State University, St-Petersburg, 198504, Russia}%

\date{May 28, 2026}

\begin{abstract}

The efficiency of optical cooling of nuclear spins in a CdTe/CdZnTe quantum well is investigated as a function of an external magnetic field. Our results confirm that there is indeed an optimal external magnetic field for optical cooling. We associate it with the kinetic local field $B_{KL}$ defined by the heating rate of the spin-spin reservoir due to the fluctuations of the hyperfine interaction. We also propose an experimental technique for measuring $B_{KL}$. For our sample we find that $B_{KL}=1.0\pm0.4$ G and it is independent of the electron polarization and pump power. The measured values of the kinetic local fields are in good agreement with a theoretical calculation $B_{KL} = 0.7$ G, taking into account indirect spin-spin interactions of Cd and Te nuclear spins and their considerably different hyperfine constants. The hyperfine constants of the magnetic isotopes of Cd and Te in CdTe are estimated. 
\end{abstract}

\maketitle

\section{\label{sec:introduction}Introduction}

    For decades, the study of nuclear spin systems (NSS) in condensed matter has remained one of the most intriguing directions in solid-state physics. The interest in this field  stems from the unique position of nuclear spins within the structure of a solid. On one hand, they are effectively isolated from the crystal lattice: the spin-lattice relaxation time $T_1$, which characterizes energy exchange with the lattice, can reach minutes or even hours at low temperatures. This allows the nuclear spins to be considered a quasi-isolated subsystem with its own internal dynamics. On the other hand, nuclear spins are coupled among themselves by magnetic dipole-dipole interactions, which ensure efficient energy exchange within the spin system itself over times on the order of the spin-spin relaxation time $T_2$\cite{Goldman, Abragam}. 

    In a wide range of solid-state systems, $T_2$ is found to be much shorter than $T_1$, which results in the establishment of the quasi-equilibrium within the spin system.  A convenient way to describe such a system is to use the concept of spin temperature \cite{Abragam}, which provides an elegant description of the NSS's behavior within the powerful framework of thermodynamics. According to the concept, the nuclei energy distribution obeys the Boltzman law with the spin temperature $\theta_N$ which can significantly differ from that of the lattice.
    From a practical standpoint, the states of the nuclear spin system  at low spin temperatures are of significant interest, as they precede the transition of the nuclear spin system into an ordered phase—such as a nuclear spin polaron or an antiferromagnet \cite{MerkulovPolaron, PolaronKavokin, PolaronGlazov}. Such a transition would enable the freezing of nuclear spin fluctuations, which constitute one of the primary sources of electron spin coherence loss in semiconductors in the dielectric phase \cite{Dzhioev2002}.
    
    An effective way of lowering the nuclear spin temperature is the optical cooling, as shown by Dyakonov and Perel in \cite{DyakPerel}. They examined the cooling efficiency of the NSS by optically oriented electrons localized on shallow donors in the regime of short correlation time of the electron spins (compared to the period of the electron precession in the random fluctuation of the nuclear field) involved in creating the nuclear spin polarization. In the steady-state case, it was shown that the inverse temperature of the nuclear spin system $\beta$ is given by
    \begin{equation}
        \beta=\left(k_b\theta_N\right)^{-1}=\frac{4 \left(\vec{S}\cdot\vec{B}\right)}{\hbar\gamma_N\left(B^2+\xi B_L^2\right)}.
        \label{eq:OptcoolingTemp}
    \end{equation}
Here $\vec{S}$ is the mean electron spin, $\vec{B}$ is the external magnetic field, $\hbar$ is the Planck constant, $\gamma_N$ is the nuclear gyromagnetic ratio, $k_b$ is the Boltzmann constant, $B_L$ is the thermodynamic nuclear local field \cite{Abragam,Goldman}. This field is determined by the internuclear spin-spin interactions and defines the heat capacity of the NSS. $B_L$ can be experimentally measured via adiabatic demagnetization of the cooled NSS, and it was done for bulk GaAs and CdTe NSS in \cite{LocalfieldGaAs, CdTeLocalField}.   We want to demonstrate that the $B_L$ and $B_{KL}$ have different physical meanings, and it is important to be able to measure them independently. The thermodynamic local field $B_L$ includes only static contributions to the heat capacity of an internal nuclear magnetic field, such as dipole-dipole, indirect, and, if present, quadrupole interactions, as well as quadrupole effects caused by the influence of the electric field of a localized electron on nearby nuclei. The contributions of these interactions to $B_L$  have been studied by us in some detail for n-GaAs \cite{Litvyak_2018,LocalfieldGaAs} and for CdTe/CdZnTe quantum well \cite{CdTeLocalField}. The kinetic local field reflects the dynamics of nuclear spins under influence of various interactions, including the hyperfine interaction with localized electrons. The dimensionless coefficient $\xi$ therefore depends also on the relative strengths of hyperfine coupling of electrons with different isotopes.

\begin{figure}
    \includegraphics[width=3.4in]{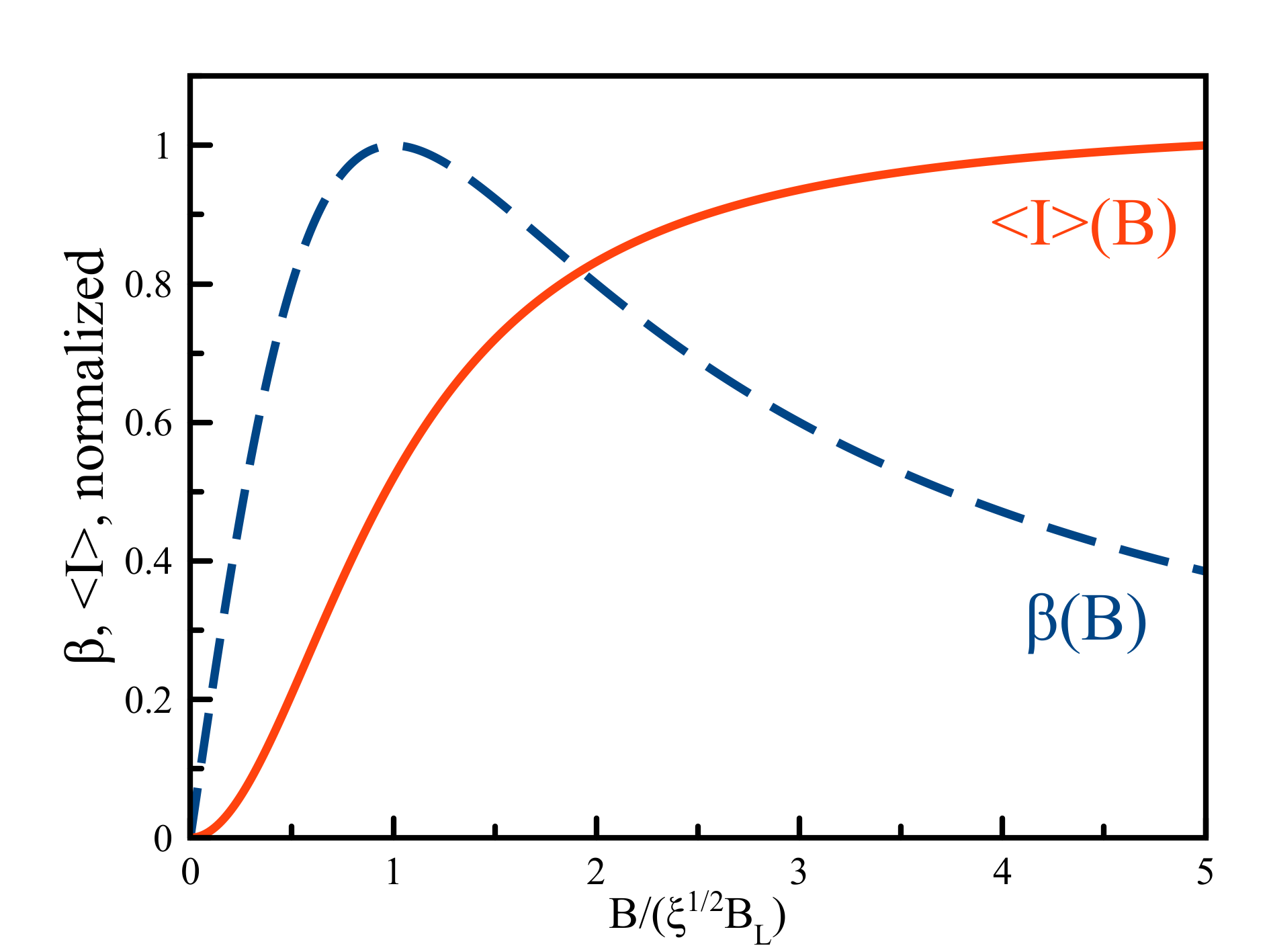}
    \caption{
    Dependence of the values of the inverse nuclear spin temperature $\beta$ (dashed blue line) and the average nuclear spin $\langle I\rangle$ (solid red line) normalized on maximum on the ratio $B/\sqrt{\xi} B_L$ for the case of short electron spin correlation times \cite{DyakPerel}.}
    \label{fig:MinTemp}
    \end{figure}

From Eq. \eqref{eq:OptcoolingTemp} it is clear that the lowest spin temperature can be achieved by cooling in the external magnetic field with value of $B = \sqrt{\xi}B_L$ (Fig. \ref{fig:MinTemp}). Physically, the term $\xi B_L^2$ in Eq. \eqref{eq:OptcoolingTemp} originates from the heating of the nuclear spin-spin reservoir induced by fluctuations in the hyperfine interaction between nuclei and the localized electron (see \cite{DyakPerel, SmirnovKavokin} or Sec. \ref{sec:discussion}). The quantity $\sqrt{\xi}B_L$, is, in fact, a measure of the internuclear field in which nuclear spins rotate and are warmed up by the components of the fluctuating hyperfine interaction resonant with this rotation.
For these reasons, we will henceforth call it the kinetic local field, $B_{KL} = \sqrt{\xi}B_L$.

Dyakonov and Perel showed that for the magnetic dipole-dipole interaction between identical nuclear spins, the parameter $\xi=3$. But it can deviate from this value if nuclear spin-spin interactions have a different form (e.g. indirect spin-spin interactions) or if the isotope composition of the NSS is complex \cite{OOChapter5}.
 A recent theoretical study \cite{SmirnovKavokin} has demonstrated that in certain semiconductor structures (e.g., quantum dots with strong electron localization), the parameter $\xi$ becomes dependent on the electron spin correlation time and can attain values significantly greater than 1, if the correlation time is long compared to the electron spin dephasing time defined by the hyperfine interaction. Consequently, the efficiency of optical cooling in systems with long correlation times is reduced compared to those with short correlation times, and the minimum spin temperature is reached in external fields that substantially exceed the local field.

  In the present work, we focus on the regime of short correlation times that was thoroughly analyzed in \cite{DyakPerel} and is briefly summarized in Sec. \ref{sec:discussion} of this work. The study investigates the process of optical cooling of the NSS in a CdTe/CdZnTe wide quantum well (QW) and examines how its efficiency depends on the external magnetic field applied along the pump beam direction. The CdTe/CdZnTe QW is selected as the object of the study due to its relatively simple nuclear spin system and the well-characterized value of the internuclear spin-spin interaction constants. This allows for more reliable theoretical estimates of the local fields. The NSS in this structure consists of 4 isotopes ($^{111}$Cd, $^{113}$Cd, $^{123}$Te, $^{125}$Te) with spin $I=1/2$ and low natural abundance ($\approx 17\: \%$). The spin value $I=1/2$ excludes any possible effects of  quadrupole interaction on measured and calculated value of the kinetic local field. 

  The approximation of the short spin correlation time is valid when the correlation time is less than the inverse rotation frequency of the electron spin in the random fluctuation of the nuclear field. 
  The electron spin rotation frequency has been experimentally determined in \cite{frequency} and equals $\omega = 1.7 \times 10^8$ rad s$^{-1}$. We conducted measurements of the spin correlation time of the electron via polarization recovery technique. The description of the method can be found for example in \cite{Dzhioev2002}. We obtained that in the range of the longitudinal fields up to 200 G there is no visible change in the PL polarization. It doesn’t allow us to determine the exact value of the electron correlation time but only to estimate it to be less than 0.5 ns. But that estimation tells us that in the sample the approximation of the short spin correlation time is reasonable.
  
  Here, we propose a technique that enables the measurement of the  kinetic local field. We also demonstrate that $B_{KL}$ depends on the constants of the direct and indirect internuclear spin-spin interactions, as well as the hyperfine constants of Cd and Te magnetic isotopes. Although the spin-spin interactions were investigated in earlier works and their constants are well-established \cite{Nolle, CdTeZULFNMR}, only a few studies have addressed the hyperfine constants in CdTe. In this work, we estimate the hyperfine constants of magnetic isotopes in CdTe and use them to calculate $B_{KL}$. The obtained value of the $B_{KL}$ is compared with the experimental data and a good agreement is observed.

\section{\label{sec:experimentalsetup}Investigated sample and experimental setup}

The experiments are performed on a single 30 nm wide $\text{CdTe/CdZn}_{0.05}\text{Te}_{0.95}$ quantum well (QW) grown by molecular beam epitaxy on a $\text{CdZn}_{0.04}\text{Te}_{0.96}$ (100) substrate. The sample's properties and the dynamics of its electron-nuclear spin system have been thoroughly investigated in previous works \cite{CdTeZULFNMR, CdTeDiffusion, CdTeLocalField}.  The sample is nominally undoped, but there are still donors on which the electrons can be localized. It makes it possible to effectively polarize NSS through strong hyperfine interactions of these electrons.

The sample is mounted in a closed-cycle cryostat and cooled to a temperature of 12 K. Optical pumping is achieved using a 671 nm laser diode. This wavelength corresponds to the electron excitation above the QW barrier. The laser beam is passed through a linear polarizer and a quarter-wave plate ($\lambda/4$) to generate polarized light with a given ellipticity. The resulting polarized photoluminescence (PL) signal passes through a photoelastic modulator (PEM) and a linear polarizer before being focused onto the entrance slit of a spectrometer. The spectrometer is set to transmit the PL intensity at a specific wavelength, which is then detected by an avalanche photodiode and a two-channel photon counter synchronized with the PEM. The measurements are performed in external magnetic fields generated by a set of magnetic coils surrounding the cryostat chamber that houses the sample. To compensate for the Earth's magnetic field, an additional set of Helmholtz coils is employed. These coils provide field cancellation in the region around the cryostat windows, reducing the residual magnetic field to approximately $1 \pm 0.5$ mG.  A schematic of the experimental setup can be found in \cite{Warm-up}.

\section{\label{sec:experimentalresults}Experimental results}

The spectra of the PL intensity and polarization of the sample under investigation can be found in \cite{CdTeDiffusion}. For subsequent experiments, a detection wavelength $\lambda=774$ nm is chosen for the polarized PL signal. At this wavelength, the optimal balance between PL signal intensity and the degree of polarization is achieved.

We conducted the experiments under three distinct conditions that varied the degree of circular polarization and the power of the pump beam:

\begin{enumerate}
    \item Circularly polarized optical pumping ($R=1$, $\rho = 6.8 \: \%$) with a pump power of 3 mW;
    \item Circularly polarized optical pumping ($R=1$, $\rho = 7.7 \%$) with a pump power of 5 mW;
    \item Elliptically polarized optical pumping ($R=0.64$, $\rho = 3.0 \: \%$) with a pump power of 3 mW;
\end{enumerate}
Here, $\rho$ is the degree of PL polarization during optical cooling. $R$ is the ellipticity degree of the pump beam, which is controlled by the rotation of a quarter-wave plate. $R=1$ corresponds to the circular polarization of the pump beam. These conditions are chosen to determine the influence of the mean electron spin on the cooling process and the $B_{KL}$.

\begin{table}[b]
\caption{\label{tab:B12}%
The HWHM of the Hanle curves for 3 experimental conditions.}
\begin{ruledtabular}
\begin{tabular}{c|c|c|c}
 &$R=1, \;\rho=7.7\%$ & $R=1, \;\rho=6.8\%$&$R=0.64, \;\rho=3.0\%$ \\
 \hline
 $P$\rule{0pt}{10pt}, mW& 5& 3&3 \\
 \hline
$B_{1/2}$, G \rule{0pt}{10pt}& $43\pm4$ & $37\pm3 $&$33\pm4$\\

\end{tabular}
\end{ruledtabular}
\end{table}

The technique employed to determine the kinetic local field in our experiments relies on their effect on the spin temperatures achieved during the optical cooling process. Rather than measuring the spin temperature $\theta_N$ directly, we measured the nuclear field $B_N$ acting on electrons (the Overhauser field), which is inversely proportional to it. A detailed description of this technique for measuring nuclear fields can be found in \cite{CdTeLocalField}.
$B_N$ is quantified by monitoring the depolarization of electron spin in a transverse magnetic field (i.e., the Hanle effect). The effective transverse field experienced by electrons comprises an externally applied "measuring" field and the induced nuclear field. To determine the effective transverse field, it is necessary to measure $B_{1/2}$ - the half-width at half-maximum  (HWHM) of the Hanle curve. Hanle width in the absence of nuclear effects is measured independently under fast polarization modulation (50 kHz) of the pump beam. The resulting  $B_{1/2}$ values are presented in Table \ref{tab:B12}. 

To determine the kinetic local field, we measure the dependence of the nuclear field $B_N$ on the field $B_z$  in which optical cooling is performed. The measurement protocol is as follows. First, optical cooling of the NSS using circularly/elliptically  polarized light directed along the z-axis is performed in a field $B_z$ for 120 seconds. After that, the field $B_z$ is switched off and a transverse measuring field $B_x = 15$ G is applied. That field induces the nuclear field $B_N$: 
\begin{equation}
    B_N=\sum_k{b_{N,k}\langle I_k\rangle}=\sum_k{b_{N,k}\frac{I\left(I+1\right)}{3}\hbar\gamma_{N,k}\beta B_x},
    \label{eq:B_n}
\end{equation}
which we can measure via Hanle effect. Here, $b_{N_k}$ is the nuclear field created by the $k$-th isotope at 100\% polarization (the constants are provided in Table \ref{tab:table2}) The field $B_z$ was varied from -15 G to 15 G.
\begin{figure}
\includegraphics[width=3.4in]{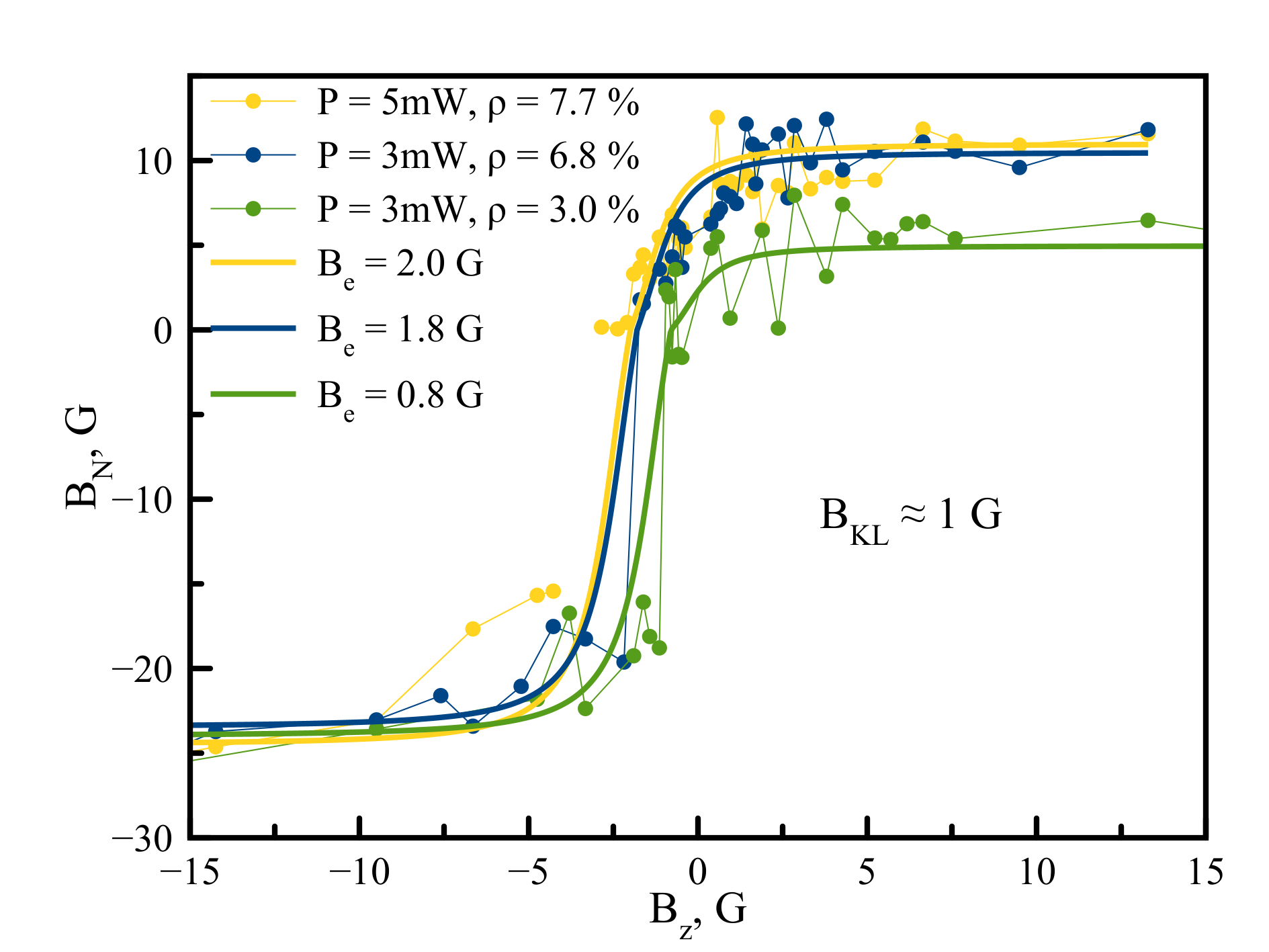}
\caption{
Dependences of the nuclear field $B_N$ on the longitudinal field $B_z$, in which optical cooling was performed, for three conditions of the optical experiment (points). The lines show theoretical fits using Eq. \eqref{eq:BnKLF}.}
\label{fig:BnKLF}
\end{figure}
The experimental data are shown in Fig. \ref{fig:BnKLF} as dots. The resulting dependencies of $B_N(B_z)$ allow one to determine $B_{KL}$.  

During optical cooling in a field $B_z$ the following spin temperature $\beta_i$ is achieved \cite{OOChapter5}:
\begin{equation}
    \beta_i=f\frac{4}{\hbar\langle\gamma_N\rangle}\frac{\left(B_z+{B}_e\right)\langle S_z\rangle}{\left(B_z+{B}_e\right)^2+B_{KL}^2}.
    \label{eq:betaZ}
\end{equation}
Here, $\langle S_z\rangle$ is the mean spin of the localized electrons, ${B}_e$ is the Knight field, and $f$ is the leakage factor, which accounts for the heating of the NSS via mechanisms other than the hyperfine interaction. After switching off the field $B_z$ and applying the field $B_x$, the inverse spin temperature changes to the value $\beta_f$ \cite{Goldman}:
\begin{equation}
    \begin{split}
    \beta_f&=\beta_i\sqrt{\frac{\left(B_z+B_e\right)^2+B_L^2}{B_x^2+B_L^2+B_e^2}}=\\=f\frac{4}{\hbar\langle\gamma_N\rangle}&\frac{\left(B_z+{B}_e\right)\langle S_z\rangle}{\left(B_z+{B}_e\right)^2+B_{KL}^2}\sqrt{\frac{\left(B_z+B_e\right)^2+B_L^2}{B_x^2+B_L^2+B_e^2}}.
    \end{split}
    \label{eq:betaX}
\end{equation}
Here $B_L=0.5$ G is the thermodynamic local field in CdTe determined in our previous work \cite{CdTeLocalField}. As a result, for the dependence of the nuclear field on the external field in which cooling occurred, we obtain:
\begin{equation}
    B_{N}=f\sum_k{b_{N_k}}\frac{\left(B_z+{B}_e\right)\langle S_z\rangle B_x}{\left(B_z+{B}_e\right)^2+B_{KL}^2}\sqrt{\frac{\left(B_z+B_e\right)^2+B_L^2}{B_x^2+B_L^2+B_e^2}}.
    \label{eq:BnKLF}
\end{equation}

In the experiment we observe that the cooling is more efficient if the mean electron spin and the magnetic field acting on nuclei are aligned, which corresponds to the positive spin temperature (negative value of $B_z$). It can be seen in Fig. \ref{fig:BnKLF}. For that reason the entire dataset cannot be adequately described by Eq. \eqref{eq:BnKLF} alone, requiring the use of different coefficients $f_+, f_-$ for positive and negative spin temperatures. We currently do not have a definitive physical explanation for why the optical cooling efficiency exhibits a dependence on the sign of the spin temperature. This remains an experimental observation that calls for further theoretical and experimental investigation.

The experimental curves are fitted using Eq. \eqref{eq:BnKLF} with  the kinetic local field $B_{KL}$, the leakage factors $f_+, f_-$ as fitting parameters. ${B}_e$ are determined from the value of the field $B_z$ at which the  nuclear field $B_N$ goes to 0 (in accordance with Eq. \ref{eq:betaZ}) and are presented in the legend of Fig. \ref{fig:BnKLF}. The fitting results are shown in Fig. \ref{fig:BnKLF} as lines. The obtained value of the kinetic local field is $B_{KL}~=~1.0\pm0.4$~G. The large error in the measurement of $B_{KL}$ is due to the difficulty of determining the nuclear field near $B_z~+~B_e~=~0$, where the spin temperature changes sign. The resulting nuclear field under these conditions is very small, making it difficult to reliably resolve its magnitude from the noise background.

\begin{figure}
\includegraphics[width=3.4in]{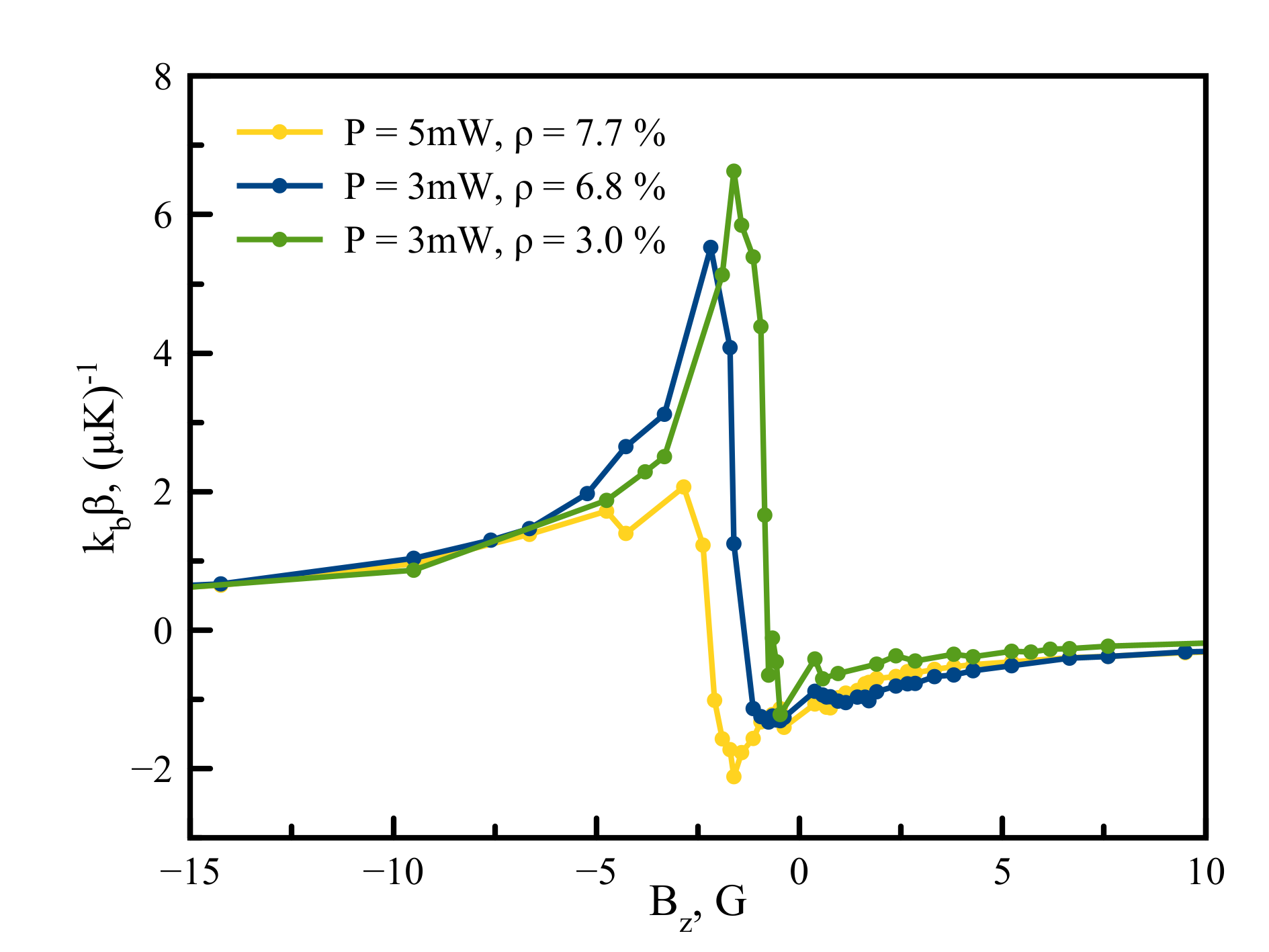}
\caption{
Inverse spin temperature of the NSS as a function of the longitudinal magnetic field, under three different optical cooling conditions. The minimum temperature of the NSS achieved here is close to 0.15 $\mu$K in field $B_z=B_{KL}-B_e$}
\label{fig:InverseTemp}
\end{figure}

To demonstrate that there indeed exists an optimal cooling magnetic field, the dependence of the inverse spin temperature on the cooling field (Eq. \eqref{eq:betaZ}) is plotted in Fig. \ref{fig:InverseTemp}. From these curves, the minimum spin temperature achieved in the experiment is estimated to be approximately $\theta_N \approx 0.15\mathrm{\mu K}$. This value is in good agreement with the one obtained in \cite{CdTeDiffusion} under similar experimental conditions.

\section{\label{sec:discussion}Discussion: The NSS cooling in the regime of the short correlation time.}

Illumination of the sample with circularly polarized light induces polarization of the electron spins. Through the hyperfine interaction, this polarization is transferred to the NSS. Since energy exchange between the NSS and the lattice is extremely slow, characteristic spin–lattice relaxation times range from several seconds to several hundred seconds—the nuclear system can establish its own temperature, which may differ significantly from that of the lattice \cite{Abragam}. The temperature that becomes established in the NSS during optical cooling can be determined by analyzing the energy flows into and out of the NSS.

In weak magnetic fields, the NSS is characterized by a single spin temperature. The value of this temperature, established during dynamic polarization by electrons, can be found from the balance of energy flows associated with the pumping of the Zeeman reservoir ($q_{pZ}$) and the heating of the Zeeman ($q_{wZ}$) and spin–spin ($q_{wSS}$) reservoirs by electrons \cite{OpticalOrientation, SmirnovKavokin}:

\begin{equation}
    q_{pZ}=q_{wZ}+q_{wSS}.
    \label{eq:energyBal}
\end{equation}
The energy flows $q_{wZ}$ and $q_{wSS}$ are caused by the heating of the NSS due to fluctuations of the Knight field. According to the fluctuation–dissipation theorem, the energy absorbed by the NSS per unit time from an alternating field is proportional to the spectral density of nuclear spin fluctuations at the field frequency and to the inverse spin temperature $\beta = (k_B \theta_N)^{-1}$ \cite{landau5}:

\begin{equation}
    q_w=q_{wZ}+q_{wSS}=\frac{\omega^2\beta\left(x^2\right)_\omega\left(S^2\right)_\omega}{8\pi}.
    \label{eq:energylflux}
\end{equation}
Here, $\left(S^2\right)_\omega$ is the spectral density of electron spin fluctuations, $\left(x^2\right)_\omega$ is the spectral density of the quantity:

\begin{equation}
    x_\alpha=\sum_n{A_n\upsilon_0\left|\Psi_e\left(\vec{r}_n\right)\right|^2I_{n,\alpha}}.
    \label{eq:magnetizationSpectralDensity}
\end{equation}
Here, $A_n$ is the hyperfine constant of nucleus $n$, $\upsilon_0$ is the volume of the primitive cell, and $\left|\Psi_e\left(\vec{r}_n\right)\right|^2$ is the squared modulus of the envelope wave function of the electron localized at a defect.

For the spectral density of the electron spin correlator in the approximation of the short spin correlation time, the following expression can be obtained:
\begin{equation}
    \left(S_\alpha^2\right)_\omega=2\langle S^2_\alpha\rangle \tau_c=2\frac{S\left(S+1\right)}{3}\tau_c=\frac{1}{2}\tau_c,
    \label{eq:spinSpectralDensity}
\end{equation}
 where $\alpha \in {x,y,z}$ denotes the Cartesian component of the vector.  

Then the total energy flow to the NSS $Q$  is determined by:
 \begin{equation}
    Q=\frac{\beta\tau_c}{16\pi}\int\limits_{-\infty}^\infty{d\omega\;\omega^2\left[\left(x_x^2\right)_\omega+\left(x_y^2\right)_\omega+\left(x_z^2\right)_\omega\right]}.
    \label{eq:fullEqnergyFlux}
\end{equation}
The integral in expression \eqref{eq:fullEqnergyFlux} can be transformed as follows:

\begin{equation}
    \int\limits_{-\infty}^\infty{d\omega\;\omega^2\left(x_\alpha^2\right)_\omega} = -\frac{2\pi}{\hbar^2}\operatorname{Tr}\left(\rho_0\left[H_0,x_\alpha\right]^2\right).
    \label{eq:integralTransform}
\end{equation}
Here, $\rho_0 = \frac{1}{2I+1}\hat{1}$ is the nuclear spin density matrix in the high temperature approximation, $H_0 = H_Z + H_{ss}$ is the Hamiltonian of the NSS, consisting of 2 parts: $H_z=-\sum_k{\hbar\gamma_{N,k}\left(\vec{I}_k\cdot\vec{B}\right)}$ is the Hamiltonian of the Zeeman interaction; $H_{ss}$ is the Hamiltonian of the internuclear spin-spin interaction. The expression for $H_{ss}$ is given in \ref{sec:trace}.  Substituting Eq. \eqref{eq:integralTransform} into Eq. \eqref{eq:fullEqnergyFlux} yields the following expression for the energy absorbed by the NSS per unit time in the high-temperature approximation:

\begin{equation}
\begin{split}
    Q=&-\frac{\beta\tau_c}{12}\sum_k{\gamma_{N,k}^2A_k^2\upsilon_0^2\left|\Psi_e\left(\vec{r}_k\right)\right|^4I_k\left(I_k+1\right)\left[B^2+B_{KL}^2\right]},\\
    B_{KL}^2=&-\frac{9\operatorname{Tr}\left(\rho_0\left[H_{ss},x_x\right]^2\right)}{2\sum_k{\hbar^2\gamma_{N,k}^2A_k^2\upsilon_0^2}\left|\Psi_e\left(\vec{r}_k\right)\right|^4I_k\left(I_k+1\right)}.
\end{split}
    \label{eq:fullEqnergyFluxKLFForm}
\end{equation}
In the calculation of $B_{KL}$, we assume that the electron density is the same for all nuclei near the donor. The hyperfine constants of the Cd and Te isotopes are estimated (see Sec.~\ref{sec:hfconstants}) and are listed in Table~\ref{tab:table2}. The detailed expression for the trace in Eq.~\ref{eq:fullEqnergyFluxKLFForm} is given in Appendix.~\ref{sec:trace}.

We obtain a local kinetic field value of $B_{KL}=0.7$ G, which agrees well with the experimentally determined value $B_{KL}=1.0\pm0.4$ G. To demonstrate the importance of using the specific hyperfine constants $A_{hf,k}$ rather than an average, we also compute $B_{KL}$ using the same average value $A_{hf}=\langle A_{hf,k}\rangle=45.1\;\mu$eV for all isotopes. This yields $B_{KL}=0.53$ G, resulting in a difference of nearly 25\% compared to the more accurate calculation.

It is instructive to compare $B_L$ and $B_{KL}$ obtained in the calculation taking into account direct and indirect nuclear spin-spin interactions and the difference in hyperfine constants of the isotope. We get $B_{KL}/B_L=\sqrt{\xi}\approx1.7$ \footnote{Surprisingly, this result is in good agreement with the one obtained in \cite{DyakPerel} $\sqrt{\xi}=\sqrt{3}$, with only direct dipole-dipole interactions included, despite the fact that the constants of the indirect scalar coupling in CdTe are larger than the direct ones and should strongly affect the values of $B_L$ and $B_{KL}$. We consider such a coincidence to be merely accidental, related to the specific values of the indirect interaction and the hyperfine constants.}. From the experiment we get $B_{KL}/B_L=2.0\pm1.1$.

\section{\label{sec:conclusion}Conclusions}

In this work, we have investigated the optical cooling of the nuclear spin system in a 30 nm CdTe/CdZnTe quantum well in the regime of short electron spin correlation times. We develop  a technique to determine the kinetic local field $B_{KL}$ of the NSS. That technique is based on the optical cooling of the NSS and the subsequent measurements of the Overhauser field via the Hanle effect.

In the experiments we determine $B_{KL} = 1.0 \pm 0.4$ of the NSS in CdTe. During the experiments the minimum nuclear spin temperature achieved is estimated to be approximately 0.15 $\mu K$. That temperature is reached after cooling of the NSS in the field $B_Z + B_e \approx B_{KL}$. It also appears that the optical cooling is more efficient in the case of the positive spin temperature. This fact doesn't follow from the known theory of the optical cooling of the NSS and requires further investigation

We also propose an analytical formula for calculation $B_{KL}$ in the regime of the short correlation time. It includes contributions from the indirect coupling  and the difference in the hyperfine constants of the nuclei. The obtained theoretical  value $B_{KL} = 0.7$ G is consistent with our experimental result. To provide more accurate calculation of $B_{KL}$ we estimated the hyperfine constants of the magnetic isotopes of Cd and Te.

\begin{acknowledgments}
The authors are grateful to M.R. Vladimirova for helpful comments on an early version of this paper. This work was supported by the Russian Science Foundation (Grant No. 25-22-00259).
\end{acknowledgments}

\appendix

\section{\label{sec:hfconstants}The Cd and Te hyperfine constants estimation}
The hyperfine constant for cadmium can be determined based on the experiments of Look and Moore \cite{LookMoore} on measuring the Knight shift of the cadmium NMR line in heavily doped n-CdTe. The Knight shift is defined as the quantity

\begin{equation}
    K=\frac{\omega_D-\omega_0}{\omega_0},
    \label{eq:KnightShift}
\end{equation}
where $\omega_D$ and $\omega_0$ are the NMR frequencies in the doped and undoped material, respectively, measured in the same magnetic field.

In a heavily doped semiconductor with a degenerate electron gas

\begin{equation}
    K=\frac{8\pi}{3}g_e\mu_B^2\upsilon_0\left|\psi\left(0\right)\right|^2\rho\left(E_F\right),
    \label{eq:KnightShiftDopedSample}
\end{equation}
where $\rho\left(E_F\right) = \frac{1}{2\hbar^2}\left(\frac{3}{\pi^4}\right)^{1/3} m_e n_e^{1/3}$ is the density of states at the Fermi level for one spin component, $g_e$ and $m_e$ are the g-factor and effective mass of the conduction band electron, respectively, $n_e$ is the electron concentration, $\upsilon_0$ is the volume of the primitive cell, and $\left|\psi\left(0\right)\right|^2$ is the electron density at the nucleus.

Thus, by measuring $K$ and $n_e$ and knowing $g_e$ one can find $\left|\psi\left(0\right)\right|^2$:

\begin{equation}
    \left|\psi\left(0\right)\right|^2=\frac{\left(9\pi\right)^{1/3}\hbar^2K}{4g_e\mu_b^2m_e\upsilon_0n_e^{1/3}}.
    \label{eq:electronDensity}
\end{equation}
Substituting the measured values $K = -4.1 \times 10^{-5}$ and $n_e = 9.7 \times 10^{17}$ cm$^{-3}$, together with the now-known parameters of CdTe: $g_e = -1.59$, $m_e = 0.0985\: m_0$, $\upsilon_0 = 6.8 \times 10^{-23}$ cm$^{3}$ into this formula, we obtain $\left|\psi\left(0\right)\right|^2 = 4.20 \times 10^{25}$ cm$^{-3}$ (previously, in Ref. \cite{NAKAMURA}, using the same data from Look and Moore but taking $g_e = -1.4$,  $\left|\psi\left(0\right)\right|^2 = 5.3 \times 10^{25}$ cm$^{-3}$ was obtained). It is interesting to compare this value $\left|\psi\left(0\right)\right|^2$ with the calculated value for the cadmium atom. According to Morton and Preston \cite{MORTON}, the Hartree–Fock value of $\left|\psi\left(0\right)\right|^2$ is $6.77 \times 10^{25}$ cm$^{-3}$. Then, this value should be multiplied by the Mackey and Wood empirical factor $F_{MW} = 1 + 3.76 \times 10^{-6} Z^3$, which accounts for the relativistic corrections for the s-shell hyperfine constants \cite{Mackey1970}. 
In case of the cadmium with $Z = 48$, we obtain $\left|\psi\left(0\right)\right|^2_{\text{at}} = 9.6 \times 10^{25}$ cm$^{-3}$ and $\frac{\left|\psi\left(0\right)\right|^2}{\left|\psi\left(0\right)\right|^2_{\text{at}}} = 0.44$.

Using $\left|\psi\left(0\right)\right|^2$ together with the gyromagnetic ratios of the cadmium isotopes, the hyperfine interaction constants can be determined
\begin{equation}
    \begin{split}
    A_{^{111}Cd}&=\frac{16\pi}{3}\mu_B\hbar\gamma_{^{111}Cd}\left|\psi\left(0\right)\right|^2\approx-24.5 \:\text{$\mu$eV},\\
    A_{^{113}Cd}&=\frac{16\pi}{3}\mu_B\hbar\gamma_{^{113}Cd}\left|\psi\left(0\right)\right|^2\approx-25.6 \:\text{$\mu$eV},
    \end{split}
    \label{eq:hyperfineConstants}
\end{equation}
which are found to be approximately half the theoretical constant for the 5s shell of the cadmium atom given by Morton and Preston ($A_{^{111}\text{Cd}} \approx -56.8 \:\mu\text{eV}$). The result is reasonable having in mind similar contributions of Cd and Te atomic s-orbitals into the conduction band Bloch amplitude in CdTe. 

Direct measurements of the tellurium hyperfine constant have not been found in the literature. To estimate it, one can use the characteristic precession frequency of donor-bound electron spins in the fields of nuclear fluctuations, measured by spin noise spectroscopy \cite{Cronenberger, frequency}. The nuclear fluctuation includes contributions from all isotopes of cadmium and tellurium:

\begin{equation}
    \delta_e=\frac{1}{\hbar}\sqrt{\frac{I\left(I+1\right)}{12\pi a_B^3}\upsilon_0\sum_\alpha{A_\alpha^2x_\alpha}}\approx 1.7 \times10^8\:\text{rad s}^{-1},
    \label{eq:localizationFreq}
\end{equation}
where $x_\alpha$ is the abundance of the isotope, and $a_B = 4.9$ nm is the Bohr radius of the donor. From these values, together with the known cadmium constants and the abundances of all isotopes, the hyperfine constant of tellurium averaged over its isotopes can be found:

\begin{equation}
    \bar{A}_{Te}=-\sqrt{\frac{1}{x_{Te}}\left(\frac{16\pi a_B^3}{\upsilon_0}\hbar^2\delta_e^2-\bar{A}^2_{Cd}\right)}\approx 109\:\text{$\mu$eV}.
    \label{eq:hyperfineConstantTe}
\end{equation}
This value is 4.3 times larger than the Cd hyperfine constant in CdTe. 
Using the gyromagnetic ratios and isotopic abundances of tellurium, the constants for each individual isotope can also be determined
\begin{equation}
\begin{split}
    A_{^{125}Te}&\approx-110.6\:\text{$\mu$eV},\\
    A_{^{123}Te}&\approx-91.9\:\:\:\text{$\mu$eV},
\end{split}
\end{equation}
and the electron density at the nucleus:

\begin{equation}
    \left|\psi_{Te}\left(0\right)\right|^2=12\cdot10^{25}\:\text{cm}^{-3}.
    \label{eq:electronDensityOnTe}
\end{equation}
The hyperfine constants were estimated in a similar manner in the preprint \cite{cronenberger2019}. However, in later works by the same authors, the value was determined more accurately. In addition, their calculations contained errors (e.g., using the GaAs unit cell volume instead of that for CdTe). As a result, the constant values listed in Table 2 of the preprint by Cronenberger et al. differ considerably from those obtained in the present work.

Knowing the electron density at the nuclei of different isotopes also allows the maximum Overhauser fields that these nuclei can exert on an electron to be determined \cite{Paget}:

\begin{equation}
    b_{N, \alpha} = \frac{16\pi}{3g_e}
\hbar x_\alpha\gamma_{N,\alpha} \left|\psi_\alpha\left(0\right)\right|^2
\label{eq:maxOverhauserField}
\end{equation}
The spin properties of all magnetic isotopes in CdTe, as well as the calculated values of the hyperfine constant and $b_N$ are listed in Table \ref{tab:table2}.

\begin{table*}[b]
\caption{\label{tab:table2}%
The parameters of the magnetic nuclei in CdTe.}
\begin{ruledtabular}
\begin{tabular}{ccccccc}
isotope & I& x, \%&$\gamma_N$ rad s$^{-1}$ G$^{-1}$& $\left|\psi\left(0\right)\right|^2, \: 10^{25}$ cm$^{-3}$& A, $\mu$eV& $b_N$, G \\
 \hline
$^{111}$Cd \rule{0pt}{10pt}& 1/2 & 12.8&$-5.698\cdot 10^3$&4.20&-24.5&170\\
\hline
$^{113}$Cd \rule{0pt}{10pt}& 1/2  & 12.2&$-5.96\cdot 10^3$&4.20&-25.6&170\\
\hline
$^{123}$Te\rule{0pt}{10pt} & 1/2  & 0.9&$-7.059\cdot 10^3$&12&-91.9&42\\
\hline
$^{125}$Te\rule{0pt}{10pt} & 1/2  & 7.0&$-8.50\cdot 10^3$&12&-110.6&397\\

\end{tabular}
\end{ruledtabular}
\end{table*}

\section{\label{sec:trace}Calculation of the trace $\operatorname{Tr}\left(\rho_0\left[H_{ss},x_x\right]^2\right)$}

The spin-spin Hamiltonian $H_{ss}$ in the expression \eqref{eq:fullEqnergyFluxKLFForm} has the following form:

\begin{equation}
\begin{split}
    {H}_{ss}=\frac{1}{2}\sum_{i\ne j}&{\frac{\hbar^2\gamma _i\gamma_j}{r_{ij}^3}\left(1+B_{ij}\right)\left( \left(\hat{\vec{I}}_i\hat{\vec{I}}_j \right)-\frac{3\left( {{{\hat{\vec{I}}}}_{i}}{{\vec{r}}_{ij}} \right)\left( {{{\hat{\vec{I}}}}_{j}}{{\vec{r}}_{ij}} \right)}{r_{ij}^{2}} \right)}+\\
    +&\frac{\hbar^2\gamma _i\gamma_j}{r_{ij}^3}A_{ij}\left(\hat{\vec{I}}_i\hat{\vec{I}}_j \right).
    \end{split}
    \label{eq:spinspinham}
\end{equation}
Here, $i$ and $j$ enumerate nuclei in the lattice, $\vec{r}_{ij}$ is the radius vector between the $i$-th and $j$-th nuclei, $\hat{\vec{I}}_i$ is the spin operator of the $i$-th nucleus, and $A_{ij}$ and $B_{ij}$ are the constants of the indirect spin-spin coupling \cite{CdTeLocalField, Nolle}. In the approximation of the constant electronic wave function on nuclei, the expression for $x_\alpha$ Eq. \eqref{eq:magnetizationSpectralDensity} can be simplified to 
\begin{equation}
    x_\alpha=\upsilon_0\left\langle|\Psi_e\left(\vec{r}\right)\right|^2\rangle\sum_n{A_nI_{n,\alpha}}.
    \label{eq:magnetizationSpectralDensitysimpl}
\end{equation}
Moreover, we assume that high-temperature approximation is still valid and  the density matrix is proportional to the identity matrix. Using expressions for the commutator of the spin components $\left[I_{n,\alpha},I_{m,\eta}\right]=\delta_{n,m}i\varepsilon_{\alpha\eta\lambda}I_{n,\lambda}$ and for the trace of the product of spin components $\operatorname{Tr}\left({I_{n,\alpha},I_{m,\eta}}\right)=\frac{I_n\left(I_n+1\right)\left(2I_n+1\right)}{3}\delta_{n,m}\delta_{\alpha,\eta}$ one can get the following expression for the $\operatorname{Tr}\left(\rho\left[H_{ss},x_x\right]^2\right)$:

\begin{equation}
    \begin{split}
    \operatorname{Tr}&\left(\rho\left[H_{ss},x_x\right]^2\right)\!\!=\Bigg\{6\sum_{F}\sum_{n\ne m}{A_F^2}\left[\frac{I_F\left(I_F+1\right)}{3}\right](E^{FF}_{n,m})^2\\
    &+\!\sum_{L\ne F}\!\sum_{n,m}\!\!\left[\frac{I_F\left(I_F+1\right)}{3}\right]\!\!\left[\frac{I_L\left(I_L+1\right)}{3}\right]\!\left(A_F-A_L\right)^2C^{LF}_{nm}\!\left(C^{LF}_{nm}-D^{LF}_{nm}\right)\\
    &-\!\sum_{L\ne F}\!\sum_{n,m}\!\!\left[\frac{I_F\left(I_F+1\right)}{3}\right]\!\!\left[\frac{I_L\left(I_L+1\right)}{3}\right]\!A_F^2 (D^{LF}_{nm})^2\Bigg\}
    \times 6\upsilon_0^2\langle|\Psi_e\left(\vec{r}\right)|^4\rangle.
    \end{split}
    \label{eq:traceEq}
\end{equation}
Here, capital indices $F$ and $L$ denote magnetic isotopes and lowercase indices $n$ and $m$ denote nuclei of the specific isotope, $A_F$ is the hyperfine constant of the $F$ isotope, and the constants $D^{FL}_{nm}, C^{FL}_{nm}, E^{FL}_{nm}$ are defined as:
\begin{equation*}
\begin{split}
    D^{FL}_{nm}&=\frac{\hbar^2\gamma_F\gamma_L}{r_{nm}^3}\left[1+B_{nm}\delta_{r_{nm},r_{cl}}\right],\\
    C^{FL}_{nm}&=\frac{\hbar^2\gamma_F\gamma_L}{r_{nm}^3}\left[1+B_{nm}\delta_{r_{nm},r_{cl}}+A_{nm}\delta_{r_{nm},r_{cl}}\right],\\
    E^{FL}_{nm}&=\frac{\hbar^2\gamma_F\gamma_L}{r_{nm}^3}.
\end{split}
\end{equation*}
$r_{cl}=\sqrt{3}/4a$ is the distance between the nearest neighbors, and $a=0.64$ nm is the lattice constant of CdTe. \bigskip

\bibliography{bibliography}

@article{CdTeDiffusion,
  title = {Nuclear spin relaxation mediated by donor-bound and free electrons in wide \text{CdTe} quantum wells},
  author = {Gribakin, B. F. and Litvyak, V. M. and Kotur, M. and Andr\'e, R. and Vladimirova, M. and Yakovlev, D. R. and Kavokin, K. V.},
  journal = {Phys. Rev. B},
  volume = {109},
  issue = {19},
  pages = {195302},
  numpages = {11},
  year = {2024},
  month = {May},
  publisher = {American Physical Society},
  doi = {10.1103/PhysRevB.109.195302},
  url = {https://link.aps.org/doi/10.1103/PhysRevB.109.195302}
}

@article{Mackey1970,
    author = {Mackey, John H. and Wood, David E.},
    title = {Empirical Correction to \uppercase{H}artree-\uppercase{F}ock-\uppercase{S}later \uppercase{S}-\uppercase{E}lectron Densities for Calculation of Contact Hyperfine Splittings},
    journal = {The Journal of Chemical Physics},
    volume = {52},
    number = {9},
    pages = {4914-4915},
    year = {1970},
    month = {05},
    issn = {0021-9606},
    doi = {10.1063/1.1673738},
    url = {https://doi.org/10.1063/1.1673738}
}

@article{Nolle,
author={Nolle, A.},
title={Direct and indirect dipole-dipole coupling between $^{111}$\text{Cd}, $^{113}$\text{Cd} and $^{125}$\text{Te} in solid \text{CdTe}},
journal={Zeitschrift f\"ur Physik B Condensed Matter},
volume={34},
number={2},
pages={175-182},
year={1979},
doi={10.1007/BF01322139},
url={https://doi.org/10.1007/BF01322139}
}

@incollection{OOChapter5,
   author    = {I. A. Merkulov and V. G. Fleisher},
   editor    = {F. Meier and B. P. Zakharchenya},
   booktitle    = {Optical Orientation},
   title     = {{Optical Orientation of the Coupled Electron-Nuclear Spin System of a Semiconductor}},
   chapter      = {5},
   year      = {1984},
   pages       = {173--258},
   publisher = {North-Holland},
   address   = {Amsterdam},
}

@article{CdTeZULFNMR,
  title = {Nuclear spin-spin interactions in \text{CdTe} probed by zero- and ultralow-field optically detected \text{NMR}},
  author = {Litvyak, V. M. and Bazhin, P. S. and Andr\'e, R. and Vladimirova, M. and Kavokin, K. V.},
  journal = {Phys. Rev. B},
  volume = {110},
  issue = {24},
  pages = {245303},
  numpages = {9},
  year = {2024},
  month = {Dec},
  publisher = {American Physical Society},
  doi = {10.1103/PhysRevB.110.245303},
  url = {https://link.aps.org/doi/10.1103/PhysRevB.110.245303}
}

@article{CdTeLocalField,
  title = {Measurement of nuclear local field by adiabatic demagnetization method for \text{CdTe/CdZnTe} quantum well},
  author = {Litvyak, V. M. and Kuznetsova, M. S. and Berdnikov, V. S. and Bazhin, P. S. and Kavokin, K. V.},
  journal = {Semiconductors},
  volume = {59},
  issue = {2},
  pages = {72-77},
  year = {2025},
  doi = {10.61011/SC.2025.02.61362.7729},
  url = {https://journals.ioffe.ru/articles/61362}
}

@article{Warm-up,
  title = {Warm-up spectroscopy of quadrupole-split nuclear spins in $n$-\text{GaAs} epitaxial layers},
  author = {Litvyak, V. M. and Cherbunin, R. V. and Kalevich, V. K. and Lihachev, A. I. and Nashchekin, A. V. and Vladimirova, M. and Kavokin, K. V.},
  journal = {Phys. Rev. B},
  volume = {104},
  issue = {23},
  pages = {235201},
  numpages = {12},
  year = {2021},
  month = {Dec},
  publisher = {American Physical Society},
  doi = {10.1103/PhysRevB.104.235201},
  url = {https://link.aps.org/doi/10.1103/PhysRevB.104.235201}
}

@BOOK{OpticalOrientation,
   author       = "M. I. Dyakonov and V. I. Perel", 
   editor       = "B. Zakharchenya and F. Meyer",
   title = "Optical Orientation",
   publisher = "North-Holland",
   address = "Amsterdam",
   year = "1984",

}

@article{SmirnovKavokin,
  title = {Cooling and Heating Nuclear Spins by Strongly Localized Electrons},
  author = {Smirnov, D. S. and Kavokin, K. V.},
  journal = {Phys. Rev. Lett.},
  volume = {134},
  issue = {1},
  pages = {016201},
  numpages = {6},
  year = {2025},
  month = {Jan},
  publisher = {American Physical Society},
  doi = {10.1103/PhysRevLett.134.016201},
  url = {https://link.aps.org/doi/10.1103/PhysRevLett.134.016201}
}

@book{landau5,
  title={Statistical Physics},
  author={Landau, L.D. and Lifshitz, E.M.},
  volume={5},
  series={Course of Theoretical Physics},
  year={1980},
  publisher={Butterworth-Heinemann},
  edition={3rd}
}

@article{LookMoore,
  title = {Nuclear-Magnetic-Resonance Measurement of the Conduction-Electron $g$ Factor in \text{CdTe}},
  author = {Look, D. C. and Moore, D. L.},
  journal = {Phys. Rev. B},
  volume = {5},
  issue = {9},
  pages = {3406--3412},
  numpages = {0},
  year = {1972},
  month = {May},
  publisher = {American Physical Society},
  doi = {10.1103/PhysRevB.5.3406},
  url = {https://link.aps.org/doi/10.1103/PhysRevB.5.3406}
}

@article{NAKAMURA,
title = {Optical detection of electron spin resonance in \text{CdTe}},
journal = {Solid State Communications},
volume = {30},
number = {7},
pages = {411-414},
year = {1979},
issn = {0038-1098},
doi = {https://doi.org/10.1016/0038-1098(79)91177-3},
url = {https://www.sciencedirect.com/science/article/pii/0038109879911773},
author = {A Nakamura and D Paget and C Hermann and C Weisbuch and G Lampel and B.C Cavenett}
}

@article{MORTON,
title = {Atomic parameters for paramagnetic resonance data},
journal = {Journal of Magnetic Resonance (1969)},
volume = {30},
number = {3},
pages = {577-582},
year = {1978},
issn = {0022-2364},
doi = {https://doi.org/10.1016/0022-2364(78)90284-6},
url = {https://www.sciencedirect.com/science/article/pii/0022236478902846},
author = {J.R Morton and K.F Preston},
}

@article{Cronenberger,
  title = {Long-range spin jump diffusion revealed by dynamic light scattering},
  author = {Cronenberger, S. and Boukari, H. and Ferrand, D. and Cibert, J. and Scalbert, D.},
  journal = {Phys. Rev. B},
  volume = {103},
  issue = {20},
  pages = {205208},
  numpages = {12},
  year = {2021},
  month = {May},
  publisher = {American Physical Society},
  doi = {10.1103/PhysRevB.103.205208},
  url = {https://link.aps.org/doi/10.1103/PhysRevB.103.205208}
}

@misc{cronenberger2019,
      title={Spatiotemporal electronic spin fluctuations in random nuclear fields in n-\text{CdTe}}, 
      author={Steeve Cronenberger and Chahine Abbas and Denis Scalbert and Hervé Boukari},
      year={2019},
      eprint={1910.11805},
      archivePrefix={arXiv},
      primaryClass={cond-mat.mes-hall},
      url={https://arxiv.org/abs/1910.11805}, 
}

@article{Paget,
  title = {Low field electron-nuclear spin coupling in gallium arsenide under optical pumping conditions},
  author = {Paget, D. and Lampel, G. and Sapoval, B. and Safarov, V. I.},
  journal = {Phys. Rev. B},
  volume = {15},
  issue = {12},
  pages = {5780--5796},
  numpages = {0},
  year = {1977},
  month = {Jun},
  publisher = {American Physical Society},
  doi = {10.1103/PhysRevB.15.5780},
  url = {https://link.aps.org/doi/10.1103/PhysRevB.15.5780}
}

@BOOK{Abragam,
   author       = {A.Abragam},
   year         = 1961,
   title        = {Principles of Nuclear Magnetism},
   publisher    = {Clarendon Press, Oxford}
}

@article{MerkulovPolaron,
    author = {Merkulov, I. A.},
    title = {Formation of a nuclear spin polaron under optical orientation in \text{GaAs}-type semiconductors},
    journal = {Physics of the Solid State},
    year = {1998},
    volume = {40},
    issue = {6},
    pages = {930-933}
}

@article{PolaronKavokin,
  title = {Electron-induced nuclear magnetic ordering in $n$-type semiconductors},
  author = {Vladimirova, M. and Scalbert, D. and Kuznetsova, M. S. and Kavokin, K. V.},
  journal = {Phys. Rev. B},
  volume = {103},
  issue = {20},
  pages = {205207},
  numpages = {11},
  year = {2021},
  month = {May},
  publisher = {American Physical Society},
  doi = {10.1103/PhysRevB.103.205207},
  url = {https://link.aps.org/doi/10.1103/PhysRevB.103.205207}
}

@article{PolaronGlazov,
  title = {Kinetic approach to nuclear-spin polaron formation},
  author = {Fischer, Andreas and Kleinjohann, Iris and Anders, Frithjof B. and Glazov, Mikhail M.},
  journal = {Phys. Rev. B},
  volume = {102},
  issue = {16},
  pages = {165309},
  numpages = {10},
  year = {2020},
  month = {Oct},
  publisher = {American Physical Society},
  doi = {10.1103/PhysRevB.102.165309},
  url = {https://link.aps.org/doi/10.1103/PhysRevB.102.165309}
}

@article{Dzhioev2002,
  title = {Low-temperature spin relaxation in n-type GaAs},
  author = {Dzhioev, R. I. and Kavokin, K. V. and Korenev, V. L. and Lazarev, M. V. and Meltser, B. Ya. and Stepanova, M. N. and Zakharchenya, B. P. and Gammon, D. and Katzer, D. S.},
  journal = {Phys. Rev. B},
  volume = {66},
  issue = {24},
  pages = {245204},
  numpages = {7},
  year = {2002},
  month = {Dec},
  publisher = {American Physical Society},
  doi = {10.1103/PhysRevB.66.245204},
  url = {https://link.aps.org/doi/10.1103/PhysRevB.66.245204}
}

@article{DyakPerel,
author = {Dyakonov, Michel and Perel, V.I.},
year = {1973},
month = {01},
pages = {362},
title = {Optical orientation in a system of electrons and lattice nuclei in semiconductors. Theory},
volume = {65},
journal = {Journal of Experimental and Theoretical Physics}
}

@article{LocalfieldGaAs,
  title = {Local field of spin-spin interactions in the nuclear spin system of $n$-GaAs},
  author = {Litvyak, V. M. and Cherbunin, R. V. and Kalevich, V. K. and Kavokin, K. V.},
  journal = {Phys. Rev. B},
  volume = {108},
  issue = {23},
  pages = {235204},
  numpages = {12},
  year = {2023},
  month = {Dec},
  publisher = {American Physical Society},
  doi = {10.1103/PhysRevB.108.235204},
  url = {https://link.aps.org/doi/10.1103/PhysRevB.108.235204}
}

@article{frequency,
  title = {Spin noise of localized electrons in a \text{CdTe}/\text{CdMgTe} quantum well},
  author = {Zibinskiy, A. L. and Cronenberger, S. and Gribakin, B. and Baye, R. and Scalbert, D. and Andr\'e, R. and Smirnov, D. S. and Vladimirova, M.},
  journal = {Phys. Rev. B},
  volume = {113},
  issue = {20},
  pages = {205418},
  numpages = {10},
  year = {2026},
  month = {May},
  publisher = {American Physical Society},
  doi = {10.1103/zc3h-1drm},
  url = {https://link.aps.org/doi/10.1103/zc3h-1drm}
}

@book{Goldman,
   author = {M. Goldman},
   title = {Spin Temperature and Nuclear Magnetic  Resonance in Solids},
   publisher = {Clarendon Press},
   address = {Oxford},
   year = {1970}
}

@article{Litvyak_2018,
doi = {10.1088/1742-6596/951/1/012006},
url = {https://doi.org/10.1088/1742-6596/951/1/012006},
year = {2018},
month = {jan},
publisher = {IOP Publishing},
volume = {951},
number = {1},
pages = {012006},
author = {Litvyak, V M and Cherbunin, R V and Kavokin, K V and Kalevich, V K},
title = {Determination of the local field in the nuclear spin system of n-type GaAs},
journal = {Journal of Physics: Conference Series},
}

\end{document}